\documentclass[12pt]{article}

\usepackage{amssymb}
\usepackage{amsmath}
\usepackage{amsfonts}

\newcommand{\Z}{\mathbb{Z}}

\newcommand{\be}{\begin{equation}}
\newcommand{\ee}{\end{equation}}
\newcommand{\bea}{\begin{eqnarray}}
\newcommand{\eea}{\end{eqnarray}}
\newcommand{\nn}{\nonumber}
\newcommand{\kt}{\rangle}

\newcommand{\ed}{\end{document}}

\newcommand{\QQ}{{\cal Q}}
\begin{document}

\title{On a $\Z_3$-Graded Generalization of the Witten Index}
\author{Ali Mostafazadeh\thanks{E-mail address:
amostafazadeh@ku.edu.tr}\\ \\
Department of Mathematics, Ko\c{c} University,\\
Rumelifeneri Yolu, 80910 Sariyer, Istanbul, Turkey}
\date{ }
\maketitle

\begin{abstract}
We construct a realization of the algebra of the $\Z_3$-graded
topological symmetry of type $(1,1,1)$ in terms of a pair of
operators $D_1:{\cal H}_1\to{\cal H}_2$, $D_2:{\cal H}_2\to{\cal
H}_3$ satisfying $[D_1D_1^\dagger,D_2^\dagger D_2]=0$. We show
that the sequence of the restriction of these operators to the
zero-energy subspace forms a complex and establish the equality
of the corresponding topological invariants with the analytic
indices of these operators.
\end{abstract}
%\vspace{2mm}
%PACS numbers: 03.65.Bz\\
%\vspace{2mm}

\baselineskip=24pt

\section{Introduction}
Since the introduction of the supersymmetric quantum mechanics,
there has been an ongoing effort to develop and study its
generalizations. Recently, we have offered in \cite{npb} a
complete characterization of a class of generalizations of
supersymmetry that share its intriguing topological properties.
These symmetries that are called topological symmetries involve
certain integer-valued topological invariants that are the
analogs of the Witten index of supersymmetry. In
Ref.~\cite{npb}, we have presented a crude description of the
topological invariants associated with a class of topological
symmetries in terms of certain operators, $D_\ell:{\cal
H}_\ell\to{\cal H}_{\ell+1}$. These operators are however subject
to a number of rather complicated compatibility conditions. The
purpose of this article is to give a nontrivial solution of these
compatibility conditions for the case of $\Z_3$-graded topological
symmetry of type $(1,1,1)$. In view of the fact that supersymmetry
coincides with the $\Z_2$-graded topological symmetry of type
$(1,1)$, $\Z_3$-graded topological symmetry of type $(1,1,1)$ is a
simplest generalization of supersymmetry.

The main motivation for the study of these symmetries is to find
out whether they lead to more general topological invariants than
the Witten index. The latter is known to be identical with the
analytic index of a Fredholm operator, \cite{witten-82}. This has
been the key observation in developing the supersymmetric proofs
of the Atiyah-Singer index theorem \cite{index}. Historically, the
fact that the analytic index of a Fredholm operator is a
topological invariant was known much earlier than the actual proof
of the index theorem. See for example \cite{palais}. Therefore, a
relevant question is whether the $\Z_3$-graded topological
symmetry of type $(1,1,1)$ leads to a topological invariant that is
more general than the analytic index of a Fredholm operator.

The organization of this article is as follows. In Section~2, we
recall the definition of the topological symmetries and review
their basic properties. In section~3, we consider the
$\Z_3$-graded topological symmetry of type $(1,1,1)$ in its
three-component representation and derive the compatibility
conditions for the operators appearing in this representation.
In Section~4, we discuss a particular nontrivial solution of 
these compatibility conditions and describe the corresponding 
topological invariants. In Section~5, we present our concluding 
remarks.

\section{Topological Symmetries}

We begin our analysis by a review of the basic properties of
topological symmetries. A more detailed discussion can be found in
\cite{npb}.

A quantum system is said to possess a $\Z_n$-graded topological
symmetry (TS) of type $(m_1,m_2,\cdots,m_n)$ iff the following
conditions are satisfied.
    \begin{itemize}
    \item[1.] The quantum system is $\Z_n$-graded. This means that
    the Hilbert space ${\cal H}$ of the quantum system is the direct
    sum of $n$ of its (nontrivial) subspaces ${\cal H}_\ell$, and
    its Hamiltonian  has a complete set of eigenvectors with definite
    {\it color} or {\it grading.} (A state is said to have a definite
    grading $c_\ell$ iff it belongs to ${\cal H}_\ell$);
    \item[2.] The  energy  spectrum is nonnegative;
    \item[3.] For every eigenvalue $E$ there corresponds a
    positive integer $\lambda_E$ such that $E$ is
    $\sum_{\ell=1}^nm_\ell\lambda_E$-fold degenerate, and the
    corresponding eigenspaces are spanned by $m_1\lambda_E$ vectors of
    grade $c_1$, $m_2\lambda_E$ vectors of grade $c_2$, $\cdots$, and
    $m_n\lambda_E$ vectors of grade $c_n$.
    \end{itemize}

For a system with a TS we can introduce a set of integer-valued
topological invariants, namely
    \be
    \Delta_{ij}:= m_i n_j^{(0)} -  m_jn_i^{(0)},
    \label{1.3}
    \end{equation}
where $i,j\in\{1,2,\cdots,n\}$ and $n_\ell^{(0)}$ denotes the number of
zero-energy states of grade $c_\ell$.

It turns out that using the above definition of TS, one can obtain
the underlying operator algebras supporting these symmetries. In
particular, $\Z_2$-graded TS of type $(1,1)$ coincides with
supersymmetry and $\Delta_{11}$ yields the Witten index.
Similarly, one obtains the algebras of parasupersymmetry of order
2 and fractional supersymmetry of arbitrary order as special cases
of TSs.

In this article we shall confine our attention to the case of
$\Z_3$-graded TS of type $(1,1,1)$. Following \cite{npb}, we grade
the state vectors by associating them with a third root of unity,
i.e., we use a grading operator satisfying
    \be
    \tau^3=1,~~~\tau^\dagger=\tau^{-1},~~~[H,\tau]=0,~~~[\tau,\QQ]_q=0,
    \label{grading}
    \end{equation}
where $\QQ$ is the generator of TS, $q:=e^{2\pi i/3}$, and
$[~,~]_q$ stands for the $q$-commutator,
$[O_1,O_2]_q:=O_1O_2-qO_2O_1$. The operator algebra for
$\Z_3$-graded TS of type $(1,1,1)$ has the form
    \bea
    &&\QQ^3=K\;,
    \label{2.3}\\
    &&Q_1^3 +M Q_1= 2^{-3/2}(K+K^\dagger)\;,
    \label{2.4}\\
    &&Q_2^3 +M=2^{-3/2}i^3(-K+K^\dagger)\;,
    \label{2.5}
    \eea
where $K$ and $M$ are operators commuting with all other
operators, $M$ is self-adjoint, and
    \be
    Q_1:=\frac{\QQ+\QQ^\dagger}{\sqrt 2},~~~~Q_2:=\frac{\QQ-\QQ^\dagger}{\sqrt 2
    \;i}.
    \label{q12}
    \end{equation}
For $K=H$, Eq.~(\ref{2.3}) coincides with the defining relation
for the fractional supersymmetry of order 3, \cite{fsusy}.

It is not difficult to show \cite{npb} that Eq.~(\ref{2.3}) implies the
particular degeneracy structure of the $\Z_3$-graded TS of type
$(1,1,1)$ provided that ${\rm ker}(K)\subseteq {\rm ker}(H)$. In
particular, a quantum system with a Hamiltonian $H$ has a 
$\Z_3$-graded TS of type $(1,1,1)$ if the following conditions
are satisfied
	\begin{itemize}
	\item[1.] There are operators $\tau$ and $\QQ$ satisfying 
	Eqs.~(\ref{grading});
	\item[2.] The spectrum of $H$ is nonnegative;
	\item[3.] Eq.~(\ref{2.3}) holds for $K=H$.
	\end{itemize}
The presence of this particular topological symmetry in turn implies
the existence of a self-adjoint operator $M$ that commutes with $\tau$
and $\QQ$ and fulfills Eqs.~(\ref{2.4}) and (\ref{2.5}).

Hereafter, we shall set $K=H$.

\section{Three-Component Realization of $\Z_3$-graded TS of type
$(1,1,1)$}

Similarly to the case of supersymmetric quantum mechanics, we can
obtain a realization of the $\Z_3$-graded TS by identifying the
state vectors with column vectors whose rows have definite grade.
We first let ${\cal H}_1,{\cal H}_2$, and ${\cal H}_3$ be three
Hilbert spaces and define the total Hilbert space ${\cal H}$ as
their inner sum, ${\cal H}:=\oplus_{\ell=1}^3{\cal H}_\ell$. Every
state vector $|\psi\kt\in{\cal H}$ can be written as the sum of
its components $|\psi\kt_\ell$ belonging to ${\cal H}_\ell$. In
the three-component representation of ${\cal H}$, this is
expressed as
    \[\psi\kt=\left(\begin{array}{c}
    |\psi_1\kt\\
    |\psi_2\kt\\
    |\psi_3\kt\end{array} \right).\]
In this representation, $\tau$ is defined by
    \be
    \tau:=\left(\begin{array}{ccc}
    q&0&0\\
    0&q^2&0\\
    0&0&1 \end{array} \right),
    \label{t}
    \end{equation}
and the operators $\QQ,H,M$ take the form
    \bea
    \QQ&=&\left(\begin{array}{ccc}
    0&0&D_3\\
    D_1&0&0\\
    0&D_2&0 \end{array} \right),
    \label{q}\\
    H&=&\left(\begin{array}{ccc}
    H_1&0&0\\
    0&H_2&0\\
    0&0&H_3 \end{array} \right),
    \label{h}\\
    M&=&\left(\begin{array}{ccc}
    M_1&0&0\\
    0&M_2&0\\
    0&0&M_3 \end{array} \right),
    \label{m}
    \eea
where $D_1:{\cal H}_1\to{\cal H}_2$, $D_2:{\cal H}_2\to{\cal
H}_3$, $D_3:{\cal H}_3\to{\cal H}_1$, $H_\ell:{\cal
H}_\ell\to{\cal H}_\ell$, and $M_\ell:{\cal H}_\ell\to{\cal
H}_\ell$, with $\ell\in\{1,2,3\}$, are linear operators, and
$H_\ell$ and $M_\ell$ are self-adjoint. Eqs.~(\ref{q}), (\ref{h}),
and (\ref{m}) are direct consequences of Eqs.~(\ref{grading}).

The condition that $H$ and $M$ commute with $Q$ is equivalent to
the statement that $(H_1,H_2,H_3)$ and $(M_1,M_2,M_3)$ are
triplets of isospectral operators, as substituting (\ref{h}),
(\ref{q}), and (\ref{m}) in $[H,Q]=0$ and $[M,Q]=0$ yields
    \bea
    && D_1H_1=H_2D_1,~~~D_2H_2=H_3D_2,~~~D_3H_3=H_1D_3,
    \label{iso-h}\\
    && D_1M_1=M_2D_1,
    \label{dm1}\\
    && D_2M_2=M_3D_2,
    \label{dm2}\\
    &&D_3M_3=M_1D_3,
    \label{dm3}
    \eea
Note that, in view of Eqs.~(\ref{h}) and (\ref{iso-h}), in order to
ensure that the spectrum of $H$ is nonnegative it is sufficient to 
show that one of $H_\ell$ has a nonnegative spectrum.

Next, we substitute (\ref{q}) and (\ref{h}) in Eq.~(\ref{2.3})
with $K=H$, i.e., set ${\QQ}^3=H$. This yields
    \be
    H_1=D_3D_2D_1,~~~~H_2=D_1D_3D_2,~~~~H_3=D_2D_1D_3.
    \label{h123}
    \end{equation}
Clearly, Eqs.~(\ref{h123}) agree with Eqs.~(\ref{iso-h}). However,
the condition that $H$ is self-adjoint implies
$H_\ell=H_\ell^\dagger$. This relation restricts the choice of the
operators $D_\ell$ and leads to the first set of compatibility
conditions for $D_\ell$, namely
    \bea
    D_1^\dagger D_2^\dagger D_3^\dagger&=&D_3D_2D_1,
    \label{c1}\\
    D_2^\dagger D_3^\dagger D_1^\dagger&=&D_1D_3D_2,
    \label{c2}\\
    D_3^\dagger D_1^\dagger D_2^\dagger&=&D_2D_1D_3.
    \label{c3}
    \eea

Next, we use Eqs.~(\ref{q}) and (\ref{q12}) to express $Q_1$ and
$Q_2$ in the three-component representation. Substituting the
result for $Q_1$ in Eq.~(\ref{2.4}) and making use of (\ref{m}),
we find
    \bea
    M_1D_3&=&D_3D_2D_2^\dagger+(D_3D_3^\dagger+D_1^\dagger D_1)D_3,
    \label{md4}\\
    M_3D_2&=&D_2D_1D_1^\dagger+(D_2D_2^\dagger+D_3^\dagger D_3)D_2,
    \label{md5}\\
    M_2D_1&=&D_2D_2D_3^\dagger+(D_1D_1^\dagger+D_2^\dagger
    D_2)D_1.
    \label{md6}
    \eea
Combining these equations with Eqs.~(\ref{dm1}) -- (\ref{dm3}), we
find the second set of compatibility conditions for $D_\ell$:
    \bea
    D_2^\dagger D_2D_1D_3&=&D_1D_3D_2D_2^\dagger,
    \label{c4}\\
    D_3^\dagger D_3D_2D_1&=&D_2D_1D_3D_3^\dagger,
    \label{c5}\\
    D_1^\dagger D_1D_3D_2&=&D_3D_2D_1D_1^\dagger.
    \label{c6}
    \eea

Finally, we substitute the matrix representation of $Q_2$ and
Eq.~(\ref{m}) in Eq.~(\ref{2.5}). The resulting matrix equation is
trivially satisfied provided that we enforce (\ref{c4}) --
(\ref{c6}). Therefore, Eq.~(\ref{2.5}) does not lead to further
conditions on $D_\ell$.

We wish to emphasize that so far we obtained a set of necessary
conditions, that is (\ref{c1}) -- (\ref{c3}) and (\ref{c4}) --
(\ref{c6}), for having a $\Z_3$-graded TS of type $(1,1,1)$. These
conditions are not sufficient as we must also make sure that the
energy spectrum is nonnegative. Furthermore, as we
mentioned above, Eq.~(\ref{2.3}) with the choice of $K=H$ is
sufficient for having the necessary degeneracy structure for the
$\Z_3$-graded TS of type $(1,1,1)$. The existence of $M$ and 
consequently $M_\ell$ follows from the definition of the TS. Hence
the conditions (\ref{c4}) -- (\ref{c5}) should be automatically 
satisfied provided that we can fulfill (\ref{c1}) -- (\ref{c3}). 
This in turns means that if we can obtain a solution of 
(\ref{c1}) -- (\ref{c3}) and show that the energy spectrum is
nonnegative, then
	\be
	\Delta_{ij}=n_j^{(0)}-n_i^{(0)}
	\label{top-inv}
	\end{equation}
are topological invariants.

In view of Eqs.~(\ref{h}) and (\ref{h123}), we also have
$n_\ell^{(0)}={\rm dim}({\rm ker}~H_\ell)$ and
    \bea
    \Delta_{12}=-\Delta_{21}&=&
    {\rm dim}({\rm ker}~D_1D_3D_2)-{\rm dim}({\rm ker}~D_3D_2D_1),
    \label{d12}\\
    \Delta_{23}=-\Delta_{32}&=&
    {\rm dim}({\rm ker}~D_2D_1D_3)-{\rm dim}({\rm ker}~D_1D_3D_2),
    \label{d23}\\
    \Delta_{13}=-\Delta_{31}&=&\Delta_{12}+\Delta_{23},
    \label{d13}
    \eea
where `dim' and `ker' abbreviate `dimension' and `kernel',
respectively.

\section{Solution of the Compatibility Conditions}

In Ref.~\cite{npb}, we gave a simple solution of the compatibility
conditions, namely
    \be
    D_2=D_1^\dagger,~~~D_3=1,
    \label{s0}
    \end{equation}
which applies to the case that ${\cal H}_3={\cal H}_1$ and $D_1$
is a Fredholm operator. One can easily check that all the
compatibility conditions are satisfied for this choice of
$D_\ell$. The corresponding independent invariants are given by
    \bea
    \Delta_{12}&=&-\Delta{23}=-\left[{\rm dim}({\rm ker}~D_1)-
    {\rm dim}({\rm ker}~D_1^\dagger)\right]=
    -\;{\rm Analytic~index}\;(D_1),\nn\\
    \Delta_{13}&=&0,\nn
    \eea
where we have made use of Eqs.~(\ref{d12}) -- (\ref{d13},
(\ref{s0}) and the fact that for any Fredholm operator $D$,
    \be
    {\rm ker}(D^\dagger D)={\rm ker}(D).
    \label{ker-dd}
    \end{equation}

In the following we construct a more interesting solution of the
compatibility conditions for $D_\ell$ which applies for arbitrary
choices of ${\cal H}_\ell$. This solution corresponds to the
choice
    \be
    D_3=(D_2D_1)^\dagger=D_1^\dagger D_2^\dagger.
    \label{d3=dd}
    \end{equation}
One can check that for this choice of $D_3$ conditions (\ref{c1})
and (\ref{c3}) are automatically fulfilled, whereas (\ref{c2})
imposes the following condition on $D_1$ and $D_2$.
    \be
    [D_1 D_1^\dagger,D_2^\dagger D_2]=0.
    \label{condi}
    \end{equation}
It turns out that this is the only condition on the operators
$D_1$ and $D_2$, as Eqs.~(\ref{c4}) -- (\ref{c5}) are trivially
satisfied provided that (\ref{condi}) holds. This confirms our
earlier remark that (\ref{c4}) -- (\ref{c5}) should follow if we
can satisfy (\ref{c1}) -- (\ref{c3}).

Furthermore, using (\ref{dm1}) and (\ref{dm2}) and setting
    \be
    M_2=D_1D_1^\dagger+D_2^\dagger D_2+D_1D_1^\dagger D_2^\dagger
    D_2,
    \label{m2=}
    \end{equation}
we can also satisfy (\ref{md4}) -- (\ref{md6}).

Equation~(\ref{d3=dd}) also implies that the Hamiltonian $H$ has a nonnegative spectrum. In order to see this we substitute Eq.~(\ref{d3=dd}) in Eqs.~(\ref{h123}). This yields
	\be
	H_1=D_3^\dagger D_3,~~~~H_2=D_1D_1^\dagger D_2^\dagger D_2,~~~~H_3=D_3 D_3^\dagger.
	\label{h123=}
	\end{equation}
Clearly these isospectral operators have a nonnegative spectrum, and the energy spectrum is nonnegative.

Next we compute the topological invariants associated with this
solution. In view of Eqs.~(\ref{ker-dd}), (\ref{condi})
and (\ref{d12}) -- (\ref{d13}), and the identity
	\[{\rm ker}(D_1D_1^\dagger D_2^\dagger D_2)=
	{\rm ker}(D_1^\dagger D_2^\dagger D_2),\]
we have
    \bea
    \Delta_{12}&=&{\rm dim}\left[{\rm ker}(D_1^\dagger D_2^\dagger D_2)\right]
	-{\rm dim}\left[{\rm ker}~D_2D_1\right],
    \label{d12-n}\\
    \Delta_{23}&=&{\rm dim}\left[{\rm ker}~(D_2D_1)^\dagger\right]-
	{\rm dim}\left[{\rm ker}(D_1^\dagger D_2^\dagger D_2)\right],
    \label{d23-n}\\
    \Delta_{13}&=&-\;{\rm Analytic~Index}\;(D_2D_1).
    \label{d13-n}
    \eea
In summary, we have so far shown that for any two operators
$D_1$ and $D_2$ satisfying condition~(\ref{condi}), the
quantities~(\ref{d12-n}) -- (\ref{d13-n}) are topological
invariants. 

Next, we study the properties of the restrictions of the operators $D_i$ to the zero-energy
subspaces 
	\be
	{\cal H}_i^{(0)}:={\rm ker}(H_i)
	\label{h=h}
	\end{equation}
of ${\cal H}_i$. First we note that according to Eqs.~(\ref{h123}) -- (\ref{c3}) and 
(\ref{h=h}),
	\bea
	{\rm ker}(D_1)\subseteq	{\rm ker}(H_1)=: {\cal H}_1^{(0)}, &&
	{\rm ker}(D_2)\subseteq	{\rm ker}(H_2)=:{\cal H}_2^{(0)},
	\label{d0}\\
	{\rm ker}(D_1^\dagger)\subseteq{\rm ker}(H_2)=: {\cal H}_2^{(0)}, &&
	{\rm ker}(D_2^\dagger)\subseteq{\rm ker}(H_3)=:{\cal H}_3^{(0)}.
	\label{d0-dagger}
	\eea
Furthermore in view of (\ref{s0}),(\ref{h123=}), and (\ref{h=h}), we can easily show that for every $|\psi_i\kt\in {\cal H}_i^{(0)}$, 
	\bea
	H_1|\psi_1\kt=D_3D_3^\dagger |\psi_1\kt=0 &\Leftrightarrow&
	D_2D_1|\psi_1\kt=D_3^\dagger|\psi_1\kt=0, 
	\label{psi-1}\\
	H_2|\psi_2\kt=D_1D_1^\dagger D_2^\dagger D_2|\psi_2\kt=0 &\Leftrightarrow&
	D_1^\dagger D_2^\dagger D_2 |\psi_2\kt=0,
	\label{psi-2}\\
	H_3|\psi_3\kt=D_3^\dagger D_3 |\psi_3\kt=0 &\Leftrightarrow&
	(D_2D_1)^\dagger|\psi_3\kt=D_3|\psi_3\kt=0.
	\label{psi-3}
	\eea
According to these identities and relations~(\ref{d0}), $D_1|\psi_1\kt\in{\rm ker}(D_2)$,
$D_2|\psi_2\kt\in{\rm ker}(D_3)$. Therefore, if we denote the restriction of $D_i$ to
${\cal H}_i^{(0)}$ by $D_i^{(0)}$, we have
	\be
	{\rm im}(D_1^{(0)})\subseteq {\rm ker}(D_2^{(0)}),~~~~
	{\rm im}(D_2^{(0)})\subseteq {\rm ker}(D_3^{(0)}),~~~~
	{\rm im}(D_3^{(0)})=\{0\},
	\label{im-in-ker}
	\end{equation}
where `im' abbreviates `image.' 
In other words, the sequence
	\be
	\{0\} \hookrightarrow {\cal H}_1^{(0)}
	\stackrel{D_1^{(0)}}{\longrightarrow}
	{\cal H}_2^{(0)}
	\stackrel{D_2^{(0)}}{\longrightarrow} 
	{\cal H}_3^{(0)}
	\stackrel{D_3^{(0)}}{\longrightarrow} \{0\}
	\label{seq}
	\end{equation}
is indeed a complex, and we can define the corresponding cohomology groups 
${\rm ker}(D_{i+1}^{(0)})/{\rm im}(D_i^{(0)})$ and Betti numbers
	\be
	b_i:={\rm dim}[{\rm ker}(D_{i+1}^{(0)})/{\rm im}(D_i^{(0)})]=
	{\rm dim}[{\rm ker}(D_{i+1}^{(0)})]-{\rm dim}[{\rm im}(D_i^{(0)})]
	\label{betti}
	\end{equation}

Next, we note that ${\cal H}_i^{(0)}$ are assumed to be finite-dimensional. This implies that
for $i\in\{1,2\}$
	\bea
	{\rm dim}[{\rm ker}(D_i^{(0)})]+ {\rm dim}[{\rm im}(D_i^{(0)})]
	&=&{\rm dim}[{\cal H}_i^{(0)}],
	\label{k+i=d}\\
	{\rm dim}[{\rm coker}(D_i^{(0)})]+{\rm dim}[{\rm im}(D_i^{(0)})]
	&=&{\rm dim}[{\cal H}_{i+1}^{(0)}],
	\label{c+i=d}
	\eea
where `coker' stands for `cokernel.' In view of Eqs.~(\ref{top-inv}), (\ref{k+i=d}) and (\ref{c+i=d}), we then obtain
	\be
	\Delta_{i,i+1}=n_{i+1}^{(0)}-n_i^{(0)}={\rm dim}[{\rm coker}(D_i^{(0)})]-
	{\rm dim}[{\rm ker}(D_i^{(0)})]={\rm dim}[{\rm ker}(D_i^{(0)\dagger})]-
	{\rm dim}[{\rm ker}(D_i^{(0)})].
	\label{delta=index-1}
	\end{equation}
Note however that because ${\rm ker}(D_i)\subseteq{\cal H}_i^{(0)}$ and ${\rm ker}(D_i^\dagger)\subseteq{\cal H}_{i+1}^{(0)}$, we have 
	\be
	{\rm ker}(D_i^{(0)})={\rm ker}(D_i),~~~~
	{\rm ker}(D_i^{(0)\dagger})={\rm ker}(D_i^\dagger).
	\label{k=k}
	\end{equation}
Combining Eqs.~(\ref{delta=index-1}) and (\ref{k=k}), we finally obtain, for $i\in\{1,2\}$,
	\be
	\Delta_{i,i+1}=-\;{\rm Analytic~Index}~(D_i).
	\label{result}
	\end{equation}
This is our main result. It indicates that for the solution~(\ref{s0}) the $\Z_3$-graded
generalization of the Witten index is nothing but the analytic index of a Fredholm operator. A 
corollary of this result is the following.
	\begin{itemize}
	\item[~] {\bf Corollary:} Let $D_i:{\cal H}_i\to{\cal H}_{i+1}$, for $i\in\{1,2\}$, be 
	Fredholm operators satisfying $[D_1D_1^\dagger,D_2^\dagger D_2]=0$. Then
		\bea
		{\rm dim}\left[{\rm ker}~D_2D_1\right]-
		{\rm dim}\left[{\rm ker}(D_1^\dagger D_2^\dagger D_2)\right]
		&=&{\rm Analytic~Index}~(D_1),
		\label{c1x}\\
		{\rm dim}\left[{\rm ker}(D_1^\dagger D_2^\dagger D_2)\right]-
		{\rm dim}\left[{\rm ker}~(D_2D_1)^\dagger\right]
		&=&{\rm Analytic~Index}~(D_2),
		\label{c2x}\\
		{\rm Analytic~Index}~(D_1)+{\rm Analytic~Index}~(D_2)
		&=&{\rm Analytic~Index}~(D_2D_1).
		\label{c3x}
		\eea
	\item[~] {\bf Proof:} Eqs.~(\ref{c1x}) and (\ref{c2x}) are direct consequences of 	Eqs.~(\ref{d12-n}), (\ref{d23-n}), (\ref{d13}), and (\ref{result}). Eq.~(\ref{c3x}) 	follows from Eqs.~(\ref{d13}), (\ref{c1x}) and (\ref{c2x}).~$\square$
	\end{itemize}

\section{Conclusion}
In this article we have explored the topological invariants of a
certain class of $\Z_3$-graded topological symmetries of type
$(1,1,1)$. The operator algebra of $\Z_3$-graded topological
symmetries of type $(1,1,1)$ together with the choice $K=H$ and the
condition of the nonnegativity of the energy spectrum imply
the presence of this topological symmetry and ensures the
topological invariance of $\Delta_{ij}$. This makes the
construction of concrete realizations of this symmetry more
tractable. Using the algebraic structure and manipulating the
resulting compatibility conditions, we found a nontrivial solution
of these conditions. We explored the meaning of the associated
topological invariants and showed that they are related to
the analytic indices of Fredholm operators. Therefore, for the
realization of the $\Z_3$-graded topological symmetries of type
$(1,1,1)$ that we considered, the generalized Witten index does
not yield a more general topological invariant. This situation is
very similar to the topological invariants of $p=2$ parasupersymmetry
(alternatively $\Z_2$-graded topological symmetry of type $(1,2)$)
studied in Ref.~\cite{ijmpa-97}. There also the known realization
of the $p=2$ parasupersymmetry does not lead to a more general
invariant.

Our investigation of the nature of the topological invariants associated
with $\Z_3$-graded topological symmetries of type
$(1,1,1)$ -- which is a graded fractional supersymmetry of order~3 ---
has also revealed the following interesting properties.
	\begin{itemize}
	\item[1.]
	The graded Hamiltonians $H_\ell$ are isospectral. This is very
	similar to the case of supersymmetry and may be useful in obtaining
	exactly solvable potentials.
	\item[2.] 
	The restriction of the square of the symmetry generator ${\cal Q}$ to the 
	zero-energy eigensubspace ${\cal H}_0$ vanishes identically. This in turn
	implies ${\rm ker}{\cal Q}^2={\cal H}_0$. 
	\end{itemize}

\section*{Acknowledgments}
I wish to thank K. Aghababaei Samani for reading the first draft of this paper and making invaluable comments. This project was supported by the Young Researcher Award Program of the Turkish Academy of Sciences (GEB$\dot{\rm I}$P).


\newpage
\begin{thebibliography}{99}
\bibitem{npb} K.~Aghababaei Samani and A.~Mostafazadeh , Nucl.\
Phys.\ B {\bf 595},467 (2001).
\bibitem{witten-82} E.~Witten, Nucl.\ Phys.\ B {\bf 202}, 253 (1982).
\bibitem{index} L.~Alvarez-Gaume, Commun.\ Math.\ Phys.\
{\bf 90}, 161 (1983);\\
L.~Alvarez-Gaume, J.\ Phys.\ A: Math.\ Gen.\ {\bf 16}, 4177 (1983);\\
P.~Windey, Acta.\ Phys.\ Pol.\ B {\bf 15}, 453 (1984);\\
A.~Mostafazadeh, J.\ Math.\ Phys.\ {\bf 35}, 1095 (1994).
\bibitem{palais} R.~S.~Palais, {\em Seminar on the Atiyah-Singer
Index Theorem,} Ann.\ of Math.\ Study, vol.~57 (Princeton
University Press, Princeton, 1965).
\bibitem{fsusy} C.~Ahn, D.~Bernard, and A.~Leclair, Nucl.\ Phys.~B {\bf 346}, 409 (1990);\\
L.~Baulieu and E.~G.~Floratos, Phys.\ Lett.~B {\bf 258}, 171 (1991);\\
R.~Kerner, J.~Math.\ Phys.\ {\bf 33}, 403 (1992);\\
S.~Durand, Phys.\ Lett.\ B {\bf 312}, 115 (1993);\\
S.~Durand, Mod.\ Phys.\ Lett.\ A {\bf 8}, 1795 (1993);\\
S.~Durand, Mod.\ Phys.\ Lett.\ A {\bf 8}, 2323 (1993);\\
A.~T.~Filippov, A.~P.~Isaev, and R.~D.~Kurdikov, Mod.\ Phys.\
Lett.~A {\bf 7},
2129 (1993);\\
N.~Mohammedi,  Mod.\ Phys.\ Lett.~A {\bf 10}, 1287 (1995);\\
N.~Fleury and M.~Rausch~de~Traubenberg, Mod.\ Phys.\ Lett.~A {\bf
11}, 2899 (1996);\\
J.~A.~de~Az\'carraga and A.~Macfarlane, J.\ Math.\ Phys. {\bf 37}, 1115 (1996);\\
R.~S.~Dunne, A.~Macfarlane, J.~A.~de~Az\'carraga, and
J.~C.~P\'erez Bueno, Int.~J.~Mod.\ Phys.\ Lett.\ A {\bf 12}, 3275
(1997);
\bibitem{ijmpa-97} A.~Mostafazadeh, Int.~J.~Mod.\ Phys.\ Lett.\ A {\bf 12}, 2725 (1997).
\end{thebibliography}
\end{document}